\documentclass[12pt,preprint]{aastex}

\usepackage{graphicx}
\usepackage{lineno}
\usepackage{amsmath, amssymb} 

\slugcomment{Submitted to Icarus}

\shorttitle{Search for Vulcanoids}
\shortauthors{Steffl et al.}

\begin{document}


\title{A Search for Vulcanoids with the STEREO Heliospheric Imager}


\author{A. J. Steffl\altaffilmark{1}, N. J. Cunningham \altaffilmark{2},
  A. B. Shinn\altaffilmark{1}, D. D. Durda\altaffilmark{1}, and
  S. A. Stern\altaffilmark{3}}


\altaffiltext{1}{Department of Space Studies, Southwest Research Institute, Boulder, CO 80302}
\altaffiltext{2}{Department of Physics, Nebraska Wesleyan University, Lincoln, NE 68504}
\altaffiltext{3}{Space Science and Engineering Division, Southwest Research Institute,
  Boulder, CO 80302}


\begin{abstract}
  Interior to the orbit of Mercury, between 0.07 and 0.21~AU, is a dynamically
  stable region where a population of asteroids, known as Vulcanoids, may
  reside. We present the results from our search for Vulcanoids using archival
  data from the Heliospheric Imager-1 (HI-1) instrument on NASA's two STEREO
  spacecraft. Four separate observers independently searched through images
  obtained from 2008-12-10 to 2009-02-28.  Roughly, all Vulcanoids with $e\leq
  0.15$ and $i \leq 15^{\circ}$ will pass through the HI-1 field of view at
  least twice during this period. No Vulcanoids were detected.  Based on the
  number of synthetic Vulcanoids added to the data that were detected, we
  derive a 3$\sigma$ upper limit (i.e. a confidence level $>0.997$) that there
  are presently no Vulcanoids larger than 5.7~km in diameter, assuming an
  R-band albedo of p$_R$=0.05 and a Mercury-like phase function. The
  present-day Vulcanoid population, if it exists at all, is likely a small
  remnant of the hypothetical primordial Vulcanoid population due to the
  combined effects of collisional evolution and subsequent radiative transport
  of collisional fragments. If we assume an extant Vulcanoid population with a
  collisional equilibrium differential size distribution with a power law
  index of -3.5, our limit implies that there are no more than 76 Vulcanoids
  larger than 1~km.
\end{abstract}



\section{Introduction}
\label{intro_section}
\linenumbers

Interior to Mercury's orbit is a dynamically stable region where a population
of small, asteroid-like bodies called Vulcanoids has long been hypothesized to
exist, cf. review by \cite{Campinsetal96}. This region, known as the Vulcanoid
zone, extends from roughly 0.07 AU to 0.21 AU (15--45 solar radii). Seen from
Earth, objects in the Vulcanoid zone have maximum solar elongation angles of
just 4--12$^\circ$.  The outer boundary of the Vulcanoid zone, at 0.21 AU, is
set by dynamical instability. Objects with semi-major axes greater than this
limit evolve onto Mercury-crossing orbits on 100-Myr timescales due to orbital
perturbations caused by both Mercury and Venus \citep{Evans:tabachnik99,
  Evans:tabachnik02}.  The inner edge of the Vulcanoid zone is less
well-defined, but is set by the combination of the intense thermal environment
and dynamical transport mechanisms such as Poynting-Robertson drag and the
Yarkovsky effect. At 0.06 AU (~13 solar radii) solar radiation is so intense
that even a pure iron body 100~km in diameter will evaporate in less than 4.5
Gyr \citep{Lebofsky75, Campinsetal96}. The time an object can survive against
evaporation is a strong function of heliocentric distance, such that, at 0.07
AU, a pure iron body of just 2~km diameter will survive for the current age of
the solar system. Poynting-Robertson drag extends the evaporation limit
outward, as it can move a 2-km diameter object with $\rho=4$~g~cm$^{-3}$ from
0.08 to 007~AU in 4~Gyr, where it will evaporate \citep{Stern:durda00}.

The detection of one or more members of the putative Vulcanoid population is
of interest as it would represent the discovery of a whole new class of solar
system objects.  If they existed at all, primordial Vulcanoids likely formed
from the highest temperature condensates near the inner edge of the solar
nebula, and they presumably would contain unique, highly refractory, chemical
signatures as a result. In addition, a primordial Vulcanoid population might
have affected Mercury's surface chronology. Based on the size-frequency
distribution of craters on the Moon, Mars, Venus, and Mercury, it is thought
that objects in the inner solar system were resurfaced during the period of
the Late Heavy Bombardment (LHB) 3.9~Gyr ago \citep{Strometal05}. Vulcanoids
removed from stable orbits in the Vulcanoid zone by non-gravitational forces
like the Yarkovsky effect \citep{Vokrouhlicky99, Vokrouhlickyetal00} could
have supplied a significant impactor population to Mercury after the LHB,
making the surface appear older \citep{Leakeetal87, Headetal07}.

\subsection{Early Searches}
\cite{Leverrier1859} first proposed that a small planet, or collection of
planets, interior to Mercury could explain the observed precession of
Mercury's orbit. This hypothesized planet was eventually given the name
``Vulcan'', and numerous searches were conducted in an attempt to find
it. However, the proximity of intramercurial objects to the Sun makes them
difficult to observe from the Earth. To overcome the observational challenge
of looking for a faint object against a bright twilight sky, many of the early
searches for the planet Vulcan were conducted in the fleeting minutes of
totality during solar eclipses. The first searches for Vulcan were done
visually, with obvious limitations. The first use of photographic plates to
search for an intra-Mercurial planet was by \cite{Perrine1902}, during the
total solar eclipse of May 18, 1901, on the island of Sumatra. After analyzing
the plates from this expedition, Perrine placed a limit of magnitude 5.0
(photographic) on the brightness of any planet interior to Mercury and
concluded that, ``...there are probably no bodies of appreciable size in the
region close about the Sun, and that the cause of the disturbance in the
motion of Mercury must be sought elsewhere.'' Subsequent observations during
solar eclipses resulted in an improved limiting magnitude of 8.0
\citep{Perrine1907, Perrine1909}.

In 1915, \cite{Einstein1915_merc} showed that the precession of Mercury's
orbit could be explained entirely by the then new theory of general
relativity, thus eliminating the dynamical need for a massive planet
Vulcan. However, the question of whether there are any small bodies interior
to Mercury's orbit remained unresolved. An archival search by
\cite{Campbell:trumpler1923} using photographic plates obtained during the
solar eclipse of 1922 for the purpose of measuring the deflection of starlight
predicted by general relativity provided the tightest constraint of all the
early searches: a photographic magnitude of 8.5.

\subsection{Modern Searches}

More recently, \cite{Courtenetal76}, summarizing a 10-year campaign to observe
comets during solar eclipses, reported, ``...data which indicate the possible
existence of one or more relatively faint objects within twenty solar
radii...and [ranging] from +9 to +7 in equivalent visual
magnitude''. Unfortunately, the nature of observing during total solar
eclipses precluded any direct follow-up observations of these possible
detections. Since these results remain unpublished and have not been
independently confirmed, it is difficult to assess whether the claimed
detections are real, and, if they are, whether the objects are Vulcanoids,
sungrazing comets (cf. \cite{Bieseckeretal02}), or some other type of body.

In contrast to all other published Vulcanoid searches, \cite{Leakeetal87}
conducted a search for Vulcanoids between 1979 and 1981 at 3.5~$\mu$m. By
virtue of operating in the thermal infrared, this search was more sensitive to
objects with low visual albedo. Leake et al. estimated a detection probability
of 75\% for an object with an L-band magnitude of 5, corresponding to an
object diameter of 40-50 km. However, bad weather and the small field of view
(FOV) of their instrument limited their search to a total of 5.8 deg$^2$
within 1$^\circ$ of the ecliptic--a small fraction of the Vulcanoid zone, as
seen from Earth.

Prior to our work, the most complete search, in terms of depth and coverage,
was conducted by \cite{Durdaetal00}, using data from the LASCO C3 coronagraph
on the SOHO spacecraft \citep{Brueckneretal95}, which images a region from
0.02--0.14 AU (3.7--30 solar radii). Durda et al., examined a 40-day sequence
of LASCO C3 images. Except for objects at the outer edge of the Vulcanoid zone
with inclination $>$25$^\circ$, all dynamically stable Vulcanoids should have
passed through the instrument FOV during this period. No Vulcanoids were
found, down to detection limit of V=8.0. For objects with a Mercury-like
albedo and phase function \citep{Veverkaetal88}, this limit corresponds to a
diameter of 22 and 65~km for objects at the inner and outer edges of the
Vulcanoid zone (or 36-106~km for an albedo of 0.05). Working independently,
\cite{Schumacher:gay01} also did not detect any Vulcanoids in LASCO C3 images
down to a limiting magnitude of V=7.

Subsequently, \cite{Durda:stern03} conducted a Vulcanoid search using a
visible wavelength imaging system flown aboard NASA F/A-18B aircraft at an
altitude of 49,000 feet. However, they were unable to improve upon the earlier
\cite{Durdaetal00} results.

\cite{Merlineetal08} report preliminary results from their search for
Vulcanoids using the Wide Angle Camera (WAC) of the MESSENGER spacecraft's
Mercury Dual Imaging System (MDIS), while the spacecraft was in cruise to
Mercury. Spacecraft pointing restrictions limited observations to those with
solar elongation $>$30$^\circ$, i.e., the outer 45\% of the Vulcanoid zone. No
Vulcanoids were detected, down to a limiting magnitude of V=8, corresponding
to an Vulcanoid diameter of 15~km. Subsequent analysis of these observations
has brought this size limit down to 5~km, comparable to the results of our
search (W.~J.~Merline, private communication, 2012).

Finally, \cite{Haibinetal09} report a Vulcanoid search using 15~cm telescopes
equipped with CCDs at two separate observatories in China during the 2008
total solar eclipse. They found ``three unidentified star-like objects'' in
images from both telescopes, but the relative motion of these objects did not
match that of a Vulcanoid. Both the angular size and sensitivity of this
search are unclear, although they report that stars as faint as V=12.8 were
detected.

\section{HI-1 Data and Processing}

For our search, we used archival data from the Heliospheric Imager (HI)
instrument on NASA’s STEREO spacecraft, available online from the STEREO
Science Center (\url{http://stereo-ssc.nascom.nasa.gov}). The Solar TErrestrial
RElations Observatory (STEREO) mission is designed to study coronal mass
ejections (CMEs) from the Sun out to the orbit of the Earth and consists of
two nearly identical spacecraft. STEREO-A orbits the Sun slightly interior to
Earth's orbit, while the other spacecraft, STEREO-B, orbits the Sun slightly
exterior to Earth's orbit \citep{Kaiseretal08}. Seen from the Sun, the angular
separation between the Earth and the two spacecraft increases by 22.5$^\circ$
per year. Our search focused on data from STEREO-A, as it has more stable
pointing than STEREO-B.

The HI instrument, itself part of the Sun Earth Connection Coronal and
Heliospheric Investigation (SECCHI), consists of two separate imagers: HI-1,
with a 20$^\circ$ square FOV centered in the ecliptic plane at a solar
elongation of 14$^\circ$, and HI-2, with a 70$^\circ$ FOV centered on the
ecliptic plane at a solar elongation of 53$^\circ$ \citep{Eylesetal09}. The HI
instruments use a series of baffles to prevent scattered light from the Sun
from reaching the detector; as a result, HI is significantly more sensitive to
faint targets like CMEs (or Vulcanoids) than traditional coronagraphic
imagers. The HI-1A imager is particularly well-suited to searching for
Vulcanoids, as all objects on dynamically stable orbits will be in the FOV at
their eastern elongation, provided the following inequality is satisfied:

\begin{equation}
\label{FOV_ineq}
a \sqrt{1-e^2} \cos i \ga r \tan \beta
\end{equation}

\noindent where $a$, $e$, and $i$ are the semi-major axis, eccentricity, and
inclination of the Vulcanoid's orbit; $r\approx0.97$~AU is the heliocentric
distance of the STEREO-A spacecraft; and $\beta=4^{\circ}$ is the angle
between the sun and the inner edge of the FOV. This inequality is satisfied
for all orbits with $a\geq 0.071$~AU, $e \leq 0.15$, and $i \leq 15^{\circ}$.

The HI-1 detector is a 2048$\times$2048 pixel CCD. To obtain sufficient
signal-to-noise (S/N) for observing faint CMEs, HI-1 typically integrates for
1200s. A single 1200s exposure would be significantly degraded by cosmic ray
hits, which affect roughly 45 pixels s$^{-1}$ \citep{Eylesetal09}. Instead,
HI-1 typically acquires 30 separate 40-second exposures. Since the data volume
required to transmit all these images to Earth at full resolution would be
prohibitive, the individual images are processed and combined onboard the
spacecraft. The image processing includes a cosmic ray removal algorithm that
compares each new image with the previous image on a pixel-by-pixel basis. If
the value in a given pixel exceeds that in the previous image by more than
5$\sigma$, where $\sigma$ is the predicted standard deviation based on the
number of photoelectrons detected, that value is replaced with the value in
the previous image. As seen from STEREO-A, the apparent motion of a Vulcanoid
between 40s exposures is $\la 5 \arcsec$. Given the 35\arcsec size of the HI-1
pixels, a full-width at half-maximum value of the PSF of 3.34 pixels in the
X-direction and 2.96 pixels in the Y-direction \citep{Bewsheretal2010}, the
rate of apparent motion for Vulcanoids, and that the signal from the solar
F-corona is much brighter than the signal expected from a Vulcanoid, this
cosmic ray removal algorithm does not affect the ability of the instrument to
detect potential Vulcanoids.

After the cosmic ray removal, the bias level, as determined from a column in
the CCD underscan region, is subtracted from the data and the images are
binned by a factor of two. The resulting sequence of 1024$\times$1024 pixel
images are then summed and compressed using a lossless algorithm, and in
general, only these processed, summed images are sent back to Earth. In normal
operations, these 1200s combined images are obtained with a cadence of 40
minutes.

\subsection{Image Processing}
\label{data_section}

\begin{figure}
\includegraphics[scale=.8]{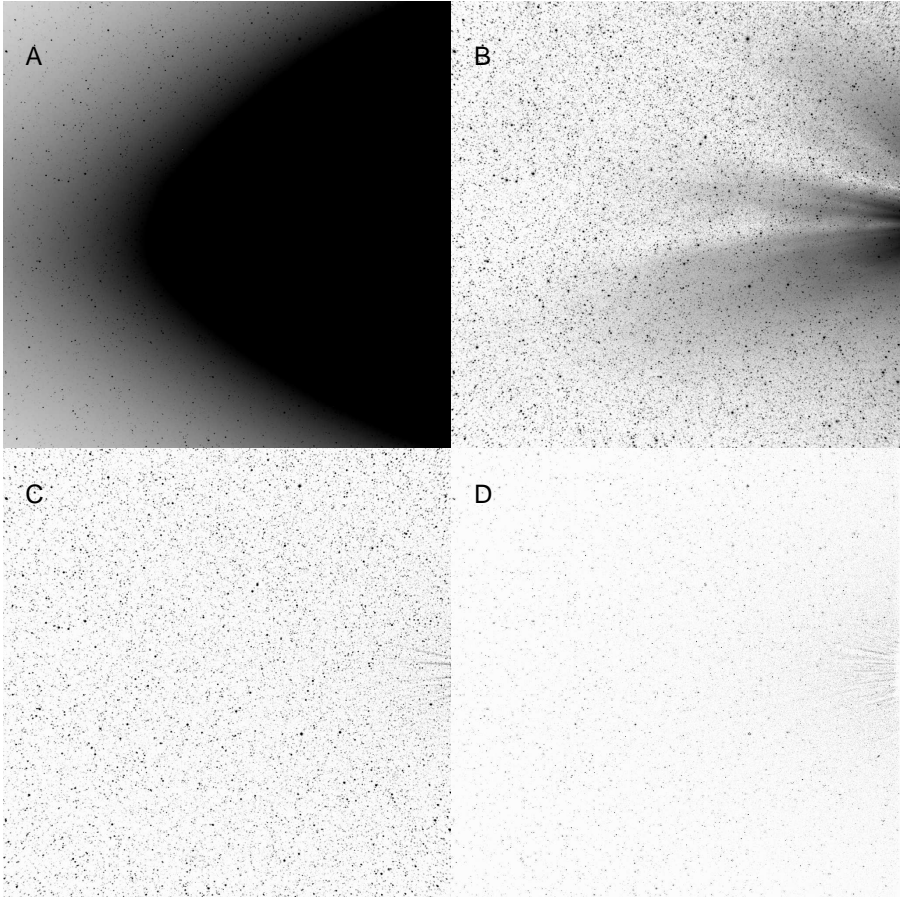}
\caption{STEREO HI-1A images at various levels of processing using the same
  stretch. Panel A shows an image after Level 1 processing. The solar F-corona
  (zodiacal light) dominates the image. Panel B shows the same image after
  Level 2 processing to remove the solar F-corona. Streamers and other
  features of solar origin are clearly visible. All point-like sources are
  stars. Panel C shows the Level 2 image in panel B after our additional sky
  subtraction; the solar features have been almost completely removed. Panel D
  shows the image in panel C after subtracting the median of the aligned
  images to suppress sources that are fixed in inertial coordinates
  (e.g. stars). See the text in Section~\ref{data_section} for more
  details. \label{hi1_images_fig}}
\end{figure}

HI data are available in three levels of processing: L0, L1, and L2. L0 data
consists of the raw, uncalibrated image data. L1 data have been processed to
correct for flat-field effects, image alignment, and the shutterless
readout. As shown in panel A of Fig.~\ref{hi1_images_fig}, both L0 and L1 data
are dominated by light scattered from the solar F-corona (zodiacal light). In
a typical 40x40 pixel region near the right-hand edge of the image, the median
value is 118~DN/s. The signal from the F-corona is fairly stable with time, so
it can be effectively removed from the data by subtracting a composite image
formed from data taken over a sufficiently long baseline. In the L2 data, the
composite image to be subtracted is created by averaging over the lowest 25\%
of pixel values obtained during either a 1- or 11-day running window at each
pixel in the image. The result of L2 processing can be seen in panel B of
Fig.~\ref{hi1_images_fig}. The same 40x40 pixel region has a median value of
0.2~DN/s in the L2 data. We used L2 data from the HI-1 instrument of STEREO-A
(henceforth HI-1A) with the 11-day running window in our search.

With the F-corona effectively removed, CMEs and coronal streamers are readily
apparent in the HI-1 data. Although the STEREO mission is designed to study
solar phenomena such as these, for the purpose of searching for Vulcanoids
they constitute an additional source of background ``noise''. If not removed,
this additional background, which is more significant at smaller angular
distances from the Sun, will result in reduced sensitivity to Vulcanoids with
small semi-major axes. We attempted to remove this two-dimensional,
time-varying background from the L2 data by subtracting off the ``sky'' level
in a circular annulus centered on each pixel. Based on an average PSF FWHM
value of 1.57 pixels (binned), we adopted radii of 4.7 and 7.9 pixels
(3$\times$ and 5$\times$ the FWHM) for our sky background annulus. This method
removes signal from diffuse features while preserving point sources. A
sky-subtracted image is shown in panel C of Fig.~\ref{hi1_images_fig}. After
this sky-subtraction, the median value in the 40x40 region fell to
8$\times10^{-3}$~DN/s.

Since the pointing of the STEREO spacecraft was held fixed relative to the
Sun, background stars appear to drift through the HI-1 FOV. To facilitate the
identification of objects moving relative to the inertial frame, we
co-registered all sky-subtracted L2 data obtained during a two-day window
(typically a series of 72 images). The first image in the sequence was used to
define an astrometric reference frame. We then calculated centroid positions
for several hundred stars common to both images and used these to derive a
third degree, two-dimensional polynomial transform that maps the second image
into the frame of the reference image. This polynomial technique handles
optical distortions near the edges of the instrument FOV better than a simple
linear shift. After the data were co-registered, we subtracted a 9-image
running median from each image to reduce the signal from any sources that
remained fixed in inertial coordinates, e.g., stars. An example of the final
processed image is shown in panel D of Fig~\ref{hi1_images_fig}.

\subsection{Synthetic Vulcanoids}
\label{synthetic_section}

To quantify the detection efficiency of our search, we generated a population
of 101 synthetic Vulcanoids and added them to the L2 HI-1A data, with the
appropriate level of Poisson noise, before our additional image
processing. The orbital properties of the synthetic Vulcanoids were chosen to
mimic those of the putative Vulcanoid population. During the search, the
observers knew that synthetic Vulcanoids were present in the data, but they
did not know where the synthetic objects were, how many objects were in a
given image (between 23 and 42 with an average of 33), or whether a given
object that was detected was real or synthetic.

\begin{figure}
\includegraphics[scale=1.]{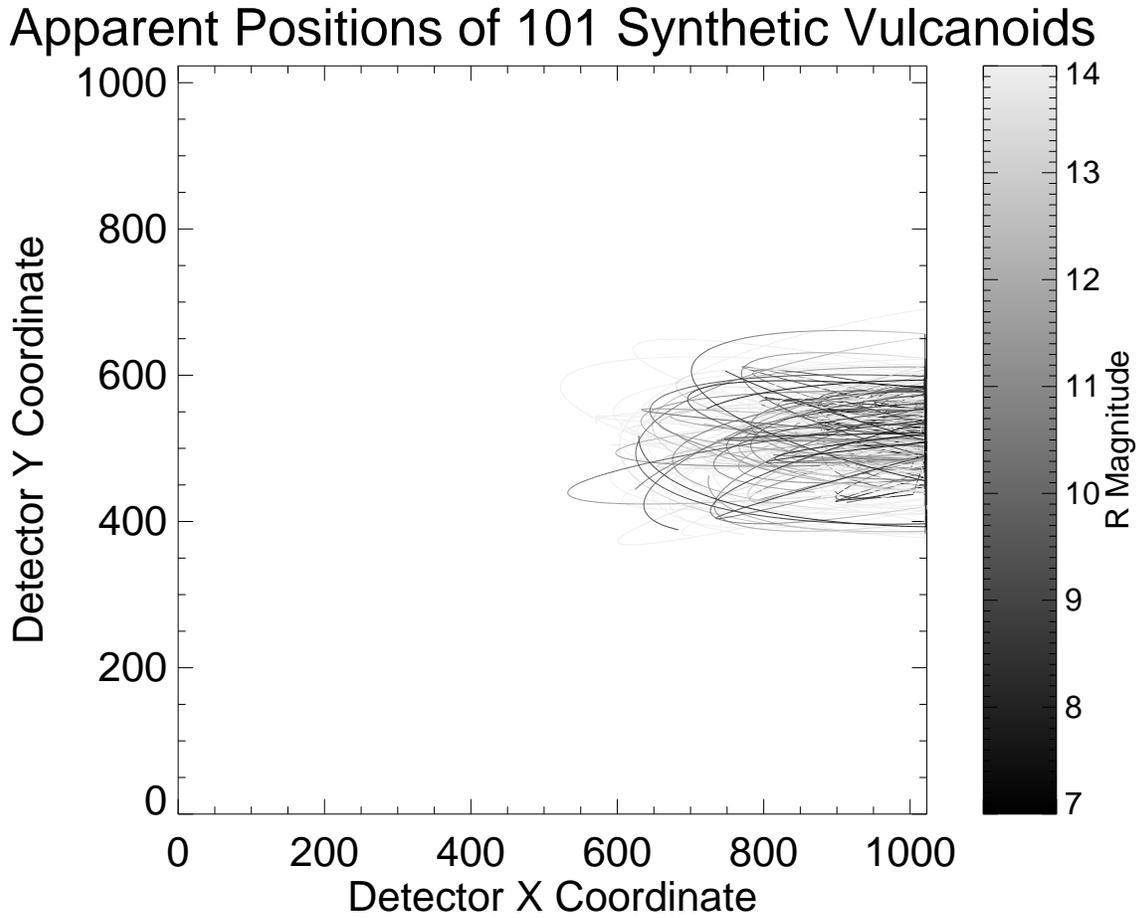}
\caption{Positions of the synthetic Vulcanoid population in HI-1A detector
  coordinates during the 40-day search period. Shades of gray indicate the
  instantaneous apparent R magnitude of the synthetic
  object. \label{orbits_fig}}
\end{figure}

Each of the 101 synthetic Vulcanoids was randomly assigned a value for the
longitude of its ascending node, argument of periapsis, and mean anomaly at
epoch.  Given the FOV of the HI-1A instrument, we limited the range of orbital
eccentricities to 0--0.15 and inclinations to 0--15$^{\circ}$, with the value
for each chosen randomly.  Prior work by \cite{Stern:durda00} has also shown
that Vulcanoids are more likely to be found on highly circular orbits near the
outer edge of the Vulcanoid zone.  For the synthetic population as a whole,
$\langle e \rangle$=0.07 and $\langle i \rangle$=7.7$^\circ$. Half of the
synthetic Vulcanoids were given a semi-major axis randomly-distributed between
0.07--0.14~A.U.; the other half were given semi-major axes distributed
randomly between 0.14--0.21~A.U. For each image to be examined, we calculated
the positions of the synthetic objects in detector coordinates, and if the
object was within the instrument field of view, we added a 2-D Gaussian PSF
(FWHM of 1.67 pixels in the X-dimension and 1.48 pixels in the Y-dimension)
with Poisson noise to the image. The positions of the synthetic Vulcanoids, in
detector coordinates, over the 40-day search period are shown in
Fig~\ref{orbits_fig}.

The apparent brightness of some Vulcanoids can change by nearly three orders
of magnitude as they move through the HI-1A field of view. This is primarily a
result of the large range of phase angles at which they can be observed
(roughly 20$^\circ$-160$^\circ$ for a Vulcanoids in the outer Vulcanoid zone
and 60$^\circ$-120$^\circ$ for Vulcanoids in the inner Vulcanoid zone) as well
as potentially significant changes in the object's heliocentric distance and
distance to the STEREO-A spacecraft while they are in the field of view. The
synthetic Vulcanoids were assumed to be spherical and within each group ($a
\le 0.14$ AU and $a > 0.14$ AU), they were assigned diameters that uniformly
spanned the range of 0.7--50~km, in logarithmically-spaced steps. The total
signal (in DN/s) from each synthetic Vulcanoid was calculated according to the
following equation:

\begin{equation}
  \label{flux_eqn}
  F=\frac{\mu A d^2}{16 r_0^2 r^2 \Delta^2 G} \int
  \left(\frac{\lambda}{hc} \right) S_{\sun}(\lambda)
  R(\lambda, \alpha) H(\lambda) d \lambda
\end{equation}

\noindent where $A$ is the collection area of HI-1A ($A=\pi d_i^2 / 4$ where
$d_i=1.59$~cm), $d$ is the diameter of the Vulcanoid in km, $r_0 = 1.496
\times 10^8$~km is the astronomical unit (AU), $r$ is the heliocentric
distance of the Vulcanoid in AU, $\Delta$ is the distance from STEREO-A to the
Vulcanoid in AU, $G$ is the system gain (15 photoelectrons / DN),
$\lambda_n/hc$ is the number of photons of wavelength $\lambda_n$ per erg,
$S_\sun$ is the solar flux at 1~AU in erg s$^{-1}$ cm$^{-2}$ nm$^{-1}$, $R$ is
the absolute reflectance of the Vulcanoid at $\lambda_n$ and phase angle
$\alpha$, and $H(\lambda_n)$ is the instrument response (which includes the
CCD quantum efficiency). Finally, $\mu = 0.93$ is the correction factor for
HI-1A between the theoretically predicted counts and counts actually measured
\citep{Bewsheretal2010}. Eq.~\ref{flux_eqn} is integrated between 300--1100
nm, the range over which HI-1 is sensitive. The instrument response curve of
HI-1A, $H(\lambda_n)$, contains a broad central peak from 600-750~nm, roughly
similar to the Johnson R filter. However, there are significant side bands at
400~nm and 950~nm (cf. Fig.~6 in Bewsher et al., 2010). For an object with a
Mercury-like spectrum, roughly 90\% of the detected flux will be in the
central pass band, 4\% in the 400~nm side band, and 6\% in the 950~nm side
band.

\begin{figure}
\plotone{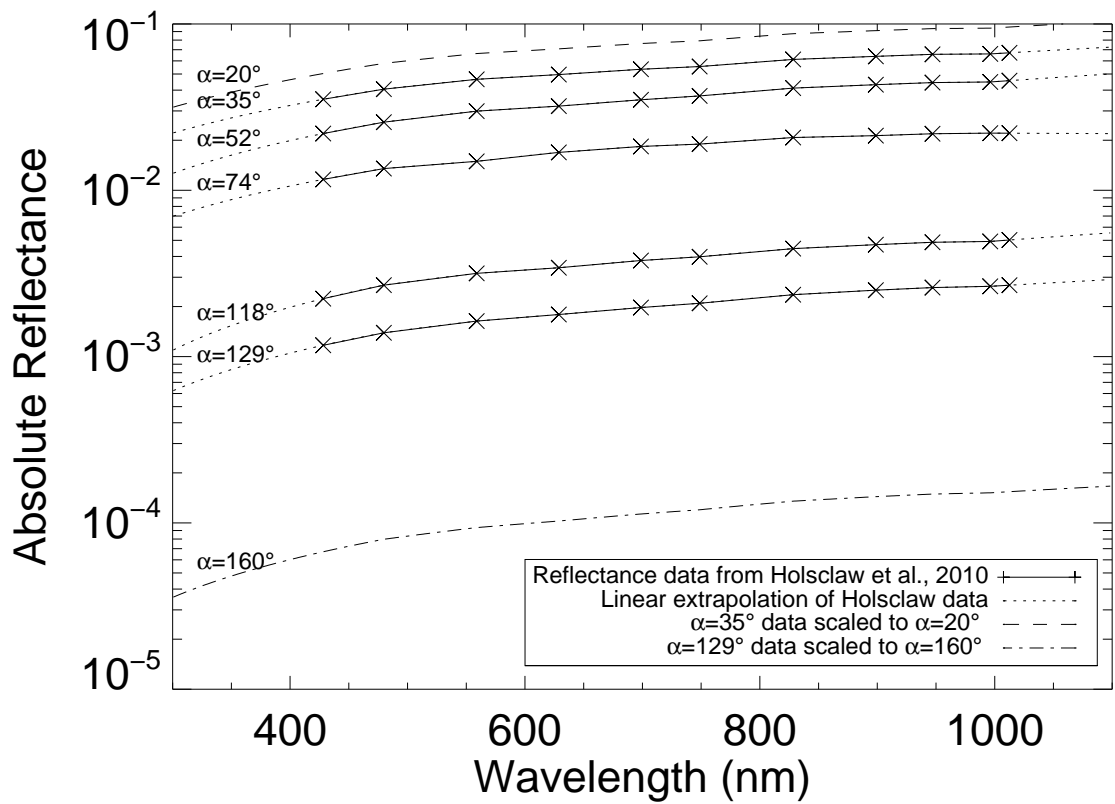}
\caption{Absolute disk-integrated reflectance of Mercury at five phase angles,
  as measured by the MASCS and MDIS instruments on MESSENEGER
  \citep{Holsclawetal2010}.  The MASCS/MDIS data have been linearly
  extrapolated to cover the full wavelength range over which the STEREO HI-1
  instrument is sensitive.  For synthetic Vulcanoids at phase angles outside
  the range (35$^\circ$--129$^\circ$) covered by the observations, we scale
  the data by the V-band phase function of \cite{Mallamaetal02}.
  \label{merc_spec_fig}}
\end{figure}

The solar spectrum, $S_{\Sun} $, was obtained from the HST CALSPEC database
\citep{Bohlin07}. The synthetic Vulcanoids were assumed to have spectral
reflectance properties identical to that of the planet Mercury, except with
1/3 the albedo, i.e., scaled to p$_R = 0.047$ \citep{Warell:bergfors08}. This
albedo is comparable to that of C-type asteroids \citep{Tholen:barucci89}. For
objects with a different albedo, $d\propto$~p$_R^{-1/2}$. The absolute
reflectance of Mercury, as measured by the MDIS and MASCS instruments on the
MESSENGER spacecraft at 11 wavelengths between 428--1013~nm at 5 phase angles
from 35$^\circ$--129$^\circ$ by \cite{Holsclawetal2010} is shown in
Fig.~\ref{merc_spec_fig}. For wavelengths between 300--428~nm and
1013--1100~nm, we linearly extrapolate the MDIS/MASCS data. For reflectances
at phase angles between 35$^\circ$--129$^\circ$, we interpolate the MDIS/MASCS
data, while for phase angles between
20$^\circ$--35$^\circ$/129$^\circ$--160$^\circ$ we scale the reflectance
spectrum at 35$^\circ$/129$^\circ$ by the V-band Mercury phase function of
\cite{Mallamaetal02}. Although we have used Mercury's reflectance spectrum in
our analysis, our results are insensitive to the choice of spectrum for
reasonable solar system alternatives, such as typical S- or C-type asteroids,
when that spectrum is scaled to an R-band albedo of 0.047.

\section{Search Technique}
\label{search_section}
After processing the data and adding synthetic Vulcanoids as described above,
we created movies from the images obtained during a two-day period. We then
looped this movie back and forth, at variable frame rates, while visually
searching for moving objects. When an object was found, the observer marked
its position in one or more frames of the movie. This technique takes
advantage of the ability of the human eye to detect motion; very often an
object that was easily identifiable while cycling through the movie would be
virtually indistinguishable from background noise when looking at an
individual image. After searching through all of the movies, the list of
marked objects for each movie was compared to the positions of the synthetic
Vulcanoids and main belt asteroids in that movie. Even with a population of
just 101 synthetic Vulcanoids, with 70\arcsec\ pixels, there were multiple
instances in which there were two synthetic Vulcanoids within a five-pixel
radius of a marked position. These were automatically flagged for further
review. Unless it was clear that both objects were marked in the same frame,
only the brighter of the two synthetic Vulcanoids was recorded as being
detected. Any marked position not matching the location of a synthetic
Vulcanoid or main belt asteroid was flagged for further inspection. With one
exception (discussed in Section~\ref{comet_section}), these were subsequently
determined to be false positives--usually the result of a interpolating a
single ``warm'' pixel or a spurious grouping of random noise.

Although there are now several years of STEREO HI-1 data, the labor-intensive
nature of our search technique placed practical limitations on the amount of
data that could be analyzed. The longest synodic period between an object in a
dynamically stable Vulcanoid orbit (i.e. having a semi-major axis $< 0.21$~AU)
and an observer at 1~AU is 39 days. We therefore selected a 40-day period from
2008-12-10 to 2009-01-19 to examine in our search, based on the low solar
activity and avoidance of the galactic center. (We had previously examined the
40-day period from 2009-01-19 to 2009-03-01 and found no Vulcanoids
\citep{Steffletal2010DPS}, but we used a less effective sky subtraction
technique resulting in a less sensitive search than the one described in this
paper). For any given 40-day period, a Vulcanoid at the outer edge of the
Vulcanoid zone will spend roughly 15 days continuously in the HI-1A field of
view during its single pass through elongation. In contrast, a Vulcanoid at
the inner edge of the Vulcanoid zone will spend roughly 30 hours in the HI-1A
FOV per elongation, but will pass through elongation 5 or 6 times.

As has been pointed out by many others before us, given a careful observer,
the human eye is often better at finding low S/N point sources with fewer
false positives than automated search algorithms. This is especially true
given the relatively dense stellar backgrounds in the HI-1 images and the
large range of apparent motions that Vulcanoids can exhibit (anywhere from no
apparent motion at all up to a few hundred arcseconds per hour, positive or
negative, in both right ascension and declination). However, using human
observers to manually look for Vulcanoids introduces a certain stochastic
element to the search. To minimize the chance that a faint Vulcanoid might be
missed due to observer fatigue or random chance, each movie was independently
examined by four separate observers (AJS, NJC, ABS, and DDD).

Finally, we note that based on this work and previous searches, e.g. the
search for small satellites of Pluto of \cite{Steffletal06b}, adding synthetic
objects to the data not only provides a quantitative means of estimating the
detection efficiency but actually results in a more sensitive search. The
tedium of searching through large numbers of images for point sources
marginally brighter than the background while failing to find any can quickly
lead to observer fatigue and a less thorough search. However, the simple
positive reinforcement of finding an object in the data, whether synthetic or
not, can help keep a human observer focused on the search, despite its
repetitive nature.

\section{Results}

No Vulcanoids were detected in our search. Although we did not detect any
Vulcanoids, we observed a variety of other solar system objects: the planets
Mercury, Venus, Uranus, and Neptune; the comet 67P/Churyumov-Gerasimenko; more
than 30 Kruetz-family sungrazing comets \citep{Bieseckeretal02}; the non-group
sungrazing comets C/2008 Y12 (SOHO) and C/2009 A1 (STEREO); and numerous
main-belt asteroids, some as faint as V=13.8, as determined by the JPL
HORIZONS ephemeris \citep{Giorginietal96}. In addition, 58 of the 101
synthetic Vulcanoids added to the data were detected by at least one of the
four observers, which we use to place upper limits on the size of any putative
Vulcanoids. Finally, as discussed below, we initially classified the
sun-grazing comet C/2008 Y12 (SOHO) as a candidate Vulcanoid. This resulted in
an additional search through all images from the HI-1 instruments on both
STEREO-A and STEREO-B over the period from 2008-12-08 to 2009-03-01. We found
no Vulcanoids in this additional search but cannot place quantitative limits
on this additional non-detection, as no synthetic Vulcanoids were added.

\subsection{Insensitivity to Orbital Parameters}

\begin{figure}
\plotone{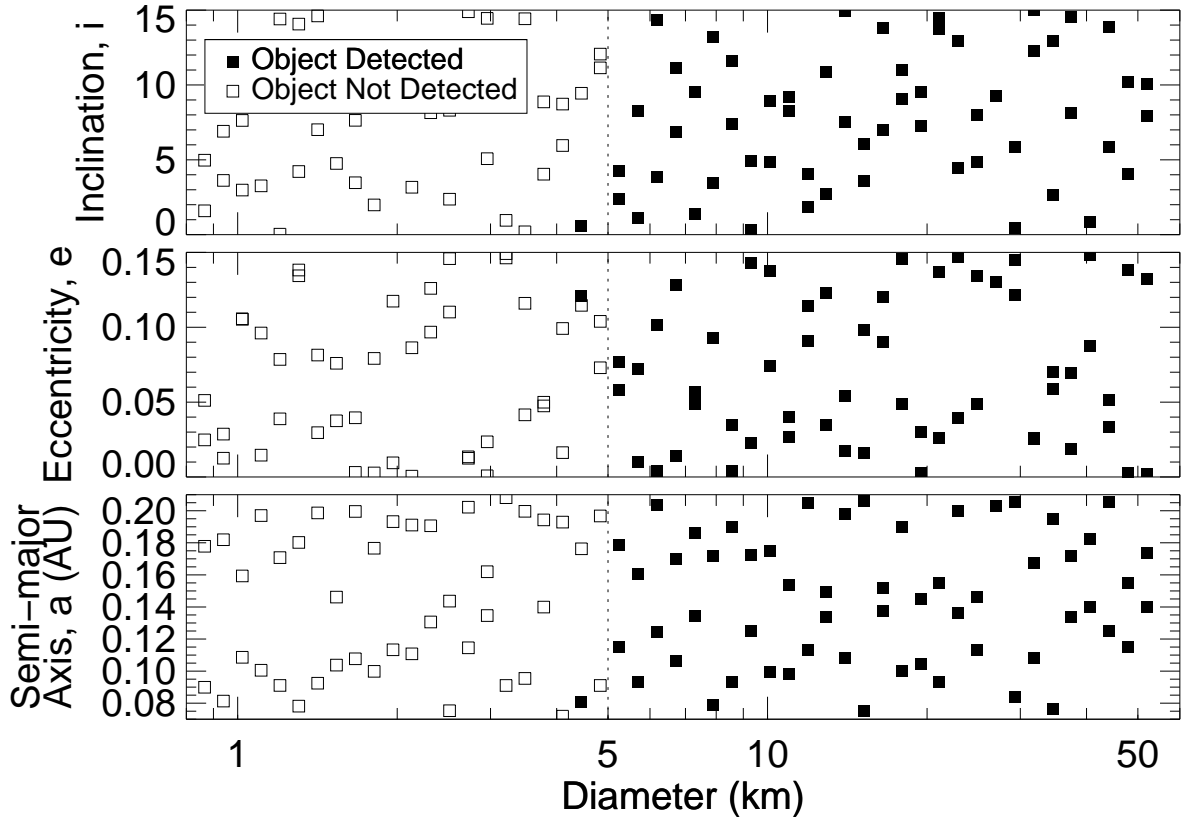}
\caption{Orbital inclination, eccentricity and semi-major axis of synthetic
  Vulcanoids, as a function of the object's diameter. Filled symbols represent
  Vulcanoids that were detected by at least one observer in our search, while
  open symbols were not detected by any of the four observers. There is no
  obvious correlation between a synthetic Vulcanoid's orbital parameters (a,
  e, and i) and whether it was detected. In contrast, all objects larger
  than 5~km were detected, while only 1 object smaller than 5~km was
  detected. \label{orbpars_vs_diam}}
\end{figure}

\begin{figure}
\plotone{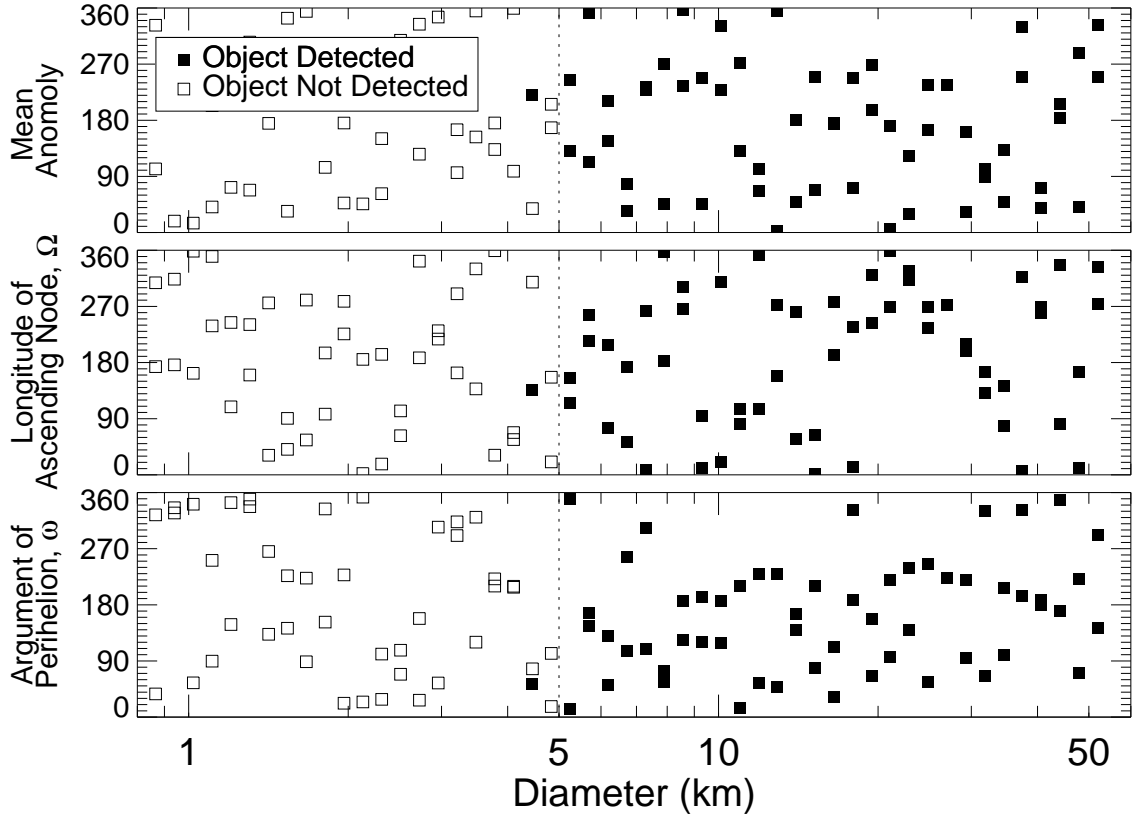}
\caption{Mean anomaly, Longitude of the ascending node, and argument of
  periheliob of synthetic Vulcanoids, as a function of the object's
  diameter. Filled symbols represent Vulcanoids that were detected by at least
  one observer in our search, while open symbols were not detected by any of
  the four observers. There is no obvious correlation between a synthetic
  Vulcanoid's orbital parameters ($\omega$, $\Omega$, and M) and whether it
  was detected. In contrast, all objects larger than 5~km were detected, while
  only 1 object smaller than 5~km was detected. \label{orbpars_vs_diam2}}
\end{figure}

The detection efficiency of Vulcanoids, whether real or synthetic, will
clearly be a strong function of the object's size, as the larger the object,
the brighter it appears. It is also possible that our search might be biased
towards detecting objects with a certain range of orbital parameter
values. For example, in our prior Vulcanoid search we found that given two
synthetic Vulcanoids of the same size, the one with the greater orbital
semi-major axis was more likely to be detected \citep{Steffletal2010DPS}. To
identify whether any such biases were present in our current search, we plot
the orbital parameters of the synthetic Vulcanoid population as a function of
diameter in Figs.~\ref{orbpars_vs_diam} and \ref{orbpars_vs_diam2}. Synthetic
Vulcanoids that were detected by at least one observer have filled symbols and
those that were not detected at all have open symbols. From these figures, it
is immediately clear that there is an extremely strong dependence on object
size, with all 57 synthetic Vulcanoids larger than 5~km in diameter being
detected by at least one observer and only 1 of the 44 synthetic Vulcanoids
smaller than 5~km in diameter being detected (and then, only by one
observer). However, there is no obvious correlation between whether an object
was detected and any of its orbital parameters.

To quantitatively verify our search's insensitivity to a Vulcanoid's orbital
parameters, we divided the synthetic Vulcanoids into two populations based on
whether they were detected or not and conducted a series of statistical tests.
These are described in more detail in the Appendix. None of these tests showed
any significant deviations from a random uniform distribution or correlations
between pairs of parameters. This is a somewhat surprising result. Evidently
the sky annulus subtraction technique described in Section~\ref{data_section},
has removed the bias towards detecting objects with larger semi-major axes
observed in our initial search. Given the relatively low number of synthetic
Vulcanoids (101), it was not possible to investigate whether any statistically
significant groupings of three or more orbital parameters exist.

\subsection{Detection Efficiency vs. Object Size}

\begin{figure}
\plotone{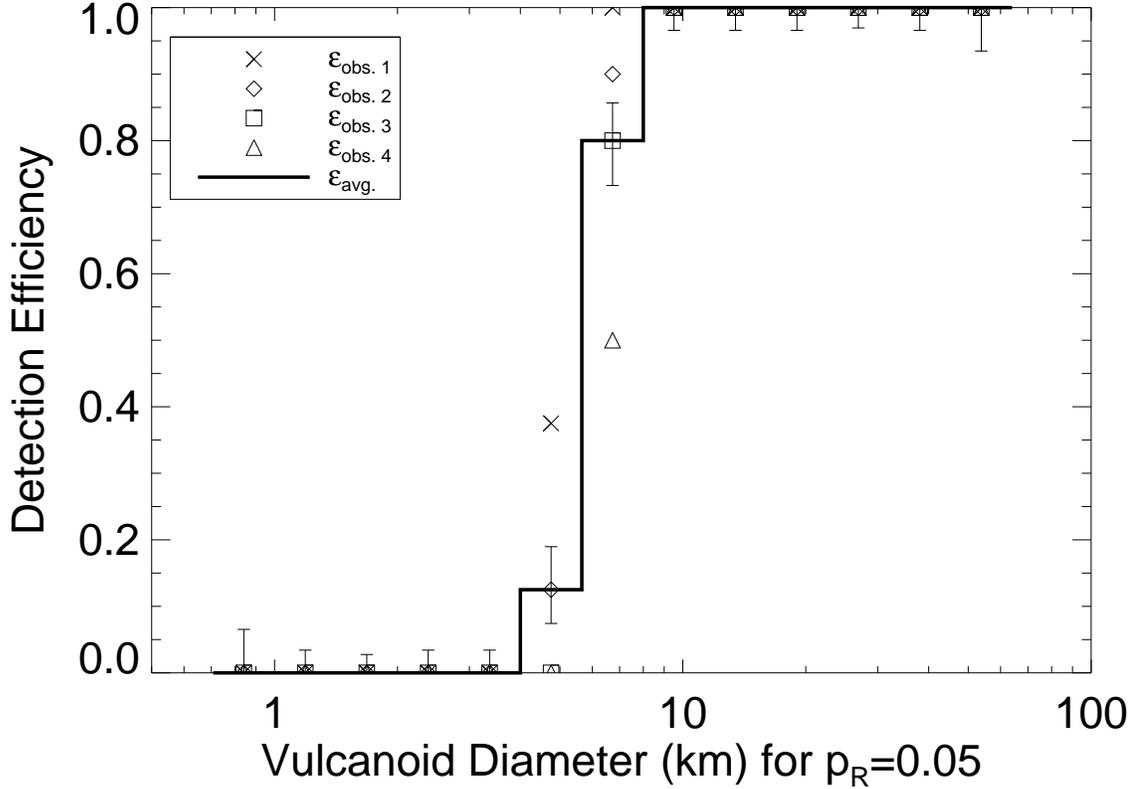}
\caption{Efficiency of detecting synthetic Vulcanoids as a function of object
  diameter. The synthetic Vulcanoids are assumed to have an R-band albedo of
  0.05 and a Mercury-like spectrum and phase function. For objects with a
  different assumed albedo, the diameter scales as $p_R^{-1/2}$. The observed
  detection efficiency for objects larger than 8~km is 1.0, while the observed
  detection efficiency for objects smaller than 4~km is 0.0. $1\sigma$ error
  bars are shown, based on the Monte Carlo analysis of the data in
  Table~\ref{deteff_table} described in the text. Given the average detection
  efficiency of 0.8 in the 5.7-8.0~km size bin, the probability of at least
  one observer detecting a real Vulcanoid of this size is
  0.9984. \label{deteff_fig}}
\end{figure}

We present the number of synthetic Vulcanoid detections for each observer,
X$_{obs}$, as a function of size in Table~\ref{deteff_table}. We have grouped
the objects into 13 size bins, corresponding to a factor of $\sqrt{2}$ in
diameter. As might be expected, there is some disparity between the detection
rates of the four observers, with observer 1 detecting 58 of the 101 synthetic
Vulcanoids and observer 4 detecting 50 of the 101 synthetic
Vulcanoids. Surprisingly, though, all the differences between the observers
occurred for objects 4-8~km diameter. All four observers detected all 45
objects larger than~8 km and none of the 38 objects smaller than 4~km. For
each observer, the detection efficiency, $\epsilon$, in a given size bin is
simply X$_{obs}$/N. We plot $\epsilon$ for each of the four individual
observers as well as $\epsilon_{avg}$, the average detection efficiency of all
observers vs.  object diameter in Fig.~\ref{deteff_fig}. 1$\sigma$ confidence
intervals for the $\epsilon_{avg}$ are derived from Monte Carlo analysis, as
described below. We find $\epsilon_{avg} = 0.800^{+0.057}_{-0.067}$ (32/40)
for objects between 5.7-8.0~km diameter and $\epsilon_{avg} =
0.125^{+0.065}_{-0.051}$ (4/32) for objects between 4.0-5.7~km. Given the
individual detection efficiencies, we calculate the probability, P$_{det}$,
that at least one of the four observers would detect a single object, as a
function of size. Subject to the assumptions stated in
Section~\ref{synthetic_section} (namely a R-band Vulcanoid albedo of
p$_R$=0.05) we find that our search should have detected all Vulcanoids larger
than 5.7~km in diameter. However this approach ignores the statistical
uncertainty in the observed values of $\epsilon$, which arises from the small
number of synthetic Vulcanoids in each size bin.

We use a Monte Carlo model to quantitatively determine the effect of
statistical uncertainty on our limits. For each size bin and observer (or the
sum of the four observers, in the case of $\epsilon_{avg}$) we set the
detection efficiency, $\epsilon$, equal to a pseudo-random number generated
over the interval [0, 1], inclusive. Using the probability mass function for
the binomial distribution, we calculated the probability of detecting exactly
X$_{obs}$ synthetic Vulcanoids, given N Vulcanoids in the size bin and
$\epsilon$. We then generated a second pseudo-random number between 0 and 1,
and if it was less than or equal to this probability the test was considered
successful at reproducing the observations and we recorded the value of
$\epsilon$. This process was repeated until 10$^6$ values for $\epsilon$ were
recorded. 1$\sigma$ confidence intervals were produced by sorting the recorded
values of $\epsilon$ into ascending order and finding the minimum contiguous
interval ($\epsilon_{high}-\epsilon_{low}$) that contained 68.3\% of the
$\epsilon$ values. These confidence intervals are shown in
Table~\ref{deteff_table}.

To determine the probability that a single Vulcanoid would be detected by at
least one observer, we randomly selected one of the recorded values for
$\epsilon$ for each observer and tested whether this value was less than or
equal to a pseudo-random number, i.e. if that observer failed to detect the
object.  If this test was true for all four observers, the Vulcanoid escaped
detection. We repeated this 10$^6$ times and counted the total number of times
the Vulcanoid was detected. The probability of a single Vulcanoid being
detected by least one observer, given statistical uncertainties,
P$^\prime_{det}$, is shown in Table~\ref{deteff_table}. We note that the
values for P$^\prime_{det}$ are significantly higher than P$_{det}$ for
Vulcanoids below 5.7~km in diameter. This is due to the fact that, given the
small values for N, even if X$_{obs}=0$, it is not possible to exclude a
non-zero value for $\epsilon$ with a high degree of confidence. For example,
observer 4 detected 0 of 8 objects between 4.0-5.7~km, but even a value of
$\epsilon=0.31$ would yield this result 5\% of the time. Thus, when at least
one observer has an observed detection efficiency, $\epsilon$, close to zero,
P$^\prime_{det}$ is likely too high. The converse is also true: when at least
one observer has $\epsilon$ close to unity, P$^\prime_{det}$ is likely an
underestimate. With this in mind, we are able to place a 3$\sigma$ upper limit
(i.e., a confidence level $>$0.997) on the existence of any Vulcanoids larger
than 5.7~km in diameter.  As stated above, our size limits are dependent on
the value of the R-band albedo, p$_R = 0.05$, we assumed for Vulcanoids. If
the actual albedo of Vulcanoids is different from this value, our diameter
limits scale as p$_{R}^{-1/2}$.

\subsection{C/2008 Y12 (SOHO)}
\label{comet_section}
In our search, we found one object that did not correspond to a synthetic
Vulcanoid or any known object in the IAU Minor Planet Center’s database
(Gareth Williams, private communication, 2011). This object was first detected
in an image from 2008-12-21T17:44 UT and appeared in 15 consecutive HI-1A
images before its apparent motion carried it outside the instrument's
FOV. Although the object's motion was similar to several to the synthetic
Vulcanoids, it appeared somewhat diffuse and brightened as it moved closer to
the sun. We thus suspected that this candidate Vulcanoid might be a previously
undetected sungrazing comet. However, given the short 9.3-hour arc and
70\arcsec\ pixels it was not possible to obtain a unique orbital solution. We
therefore began a concerted effort to see if the object could be re-acquired
in images from the HI-1 instruments on either STEREO-A or STEREO-B obtained
over the 83-day period from 2008-12-08 to 2009-03-01. These data were
processed and searched using the techniques described above, but without the
addition of any synthetic Vulcanoids. Ultimately, we were not able to
reacquire the object and no other candidate Vulcanoids were detected. Since we
were specifically looking to reacquire the candidate object in the data, this
secondary search was likely not as sensitive to Vulcanoids in general as our
primary search. However, lacking any synthetic Vulcanoids to detect, we can
not quantitatively verify this hypothesis. Given that in our primary search,
all four observers detected all 45 synthetic objects larger than 8~km, it
seems likely that were such an object present in this extended data set, it
would have been detected. This gives us additional confidence that some
unknown systematic effect unique to either the HI-1A or the time period from
2008-12-10 to 2009-01-19 is not preventing the detection of a real Vulcanoid.

Subsequently, we learned of the comet C/2008 Y12 (SOHO), which was discovered
in 7 images from the LASCO C2 coronagraph on the SOHO spacecraft, starting on
2008-12-22T16:54--less than 14 hours after the Vulcanoid candidate was last
seen in the HI-1A images. The predicted positions of C/2008 Y12 (SOHO) in the
HI-1A images using the initial orbital fit \citep{MPEC2009F17} were not a good
match to our candidate object. However, we were able to find a single set of
orbital parameters that yielded average positional residuals of 37\arcsec\ for
the Vulcanoid candidate in the STEREO HI-1A images and 3\arcsec\ for C/2008
Y12 (SOHO) in the SOHO LASCO-C2 images--less than half the size of the
respective detector pixels. After reporting these results to the IAU Minor
Planet Center, a new orbital fit was published \cite{MPEC2012B23}. However,
this published fit assumed an eccentricity of 1.0, and as a result, it is
significantly less accurate than our orbital fit shown in
Table~\ref{elements_table}.

\section{Discussion}

Although it is dynamically stable, the Vulcanoid zone is in a rough
neighborhood. It occupies a comparatively small volume of space and due to its
proximity to the sun, orbital velocities are large.  \cite{Leakeetal87} and
\cite{Stern:durda00} showed that even for a wide variety of assumptions about
the size and material properties of a hypothetical population, Vulcanoids will
experience significant collisional disruption and erosion unless the mean
orbital eccentricity of the population is extremely small
($\sim10^{-3}$). However, dynamical studies by \cite{Evans:tabachnik02} showed
that test particles between 0.09 and 0.20~AU on initially circular orbits with
zero inclination evolved dynamically such that, after just 100~Myr, the
population had a mean eccentricity of $\langle e \rangle = 0.0935$ and a mean
maximum eccentricity over the 100~Myr of $\langle e_{max} \rangle =
0.15$. This increase in mean eccentricity is due to the interaction of various
orbital resonances with Mercury as its orbit evolved over the simulation.  For
a Vulcanoid population with mean eccentricity of $\langle e \rangle = 0.0935$,
typical encounter velocities between Vulcanoids will be in the range of
10-20~km~s$^{-1}$ and collisional disruption/erosion is extremely
efficient. For an initial Vulcanoid population with a power-law index of -2.5,
300 objects larger than 1~km radius, and $\langle e \rangle = 0.1024$,
high-energy collisions with small bodies reduced all objects to debris, no
larger than 1~km radius, in 1.2~Gyr \citep{Stern:durda00}. For an initial
population of 10$^4$ objects with r$>1$~km, this collisional destruction
timescale was just 1.6~Myr.

However, radiative transport mechanisms like the Yarkovsky effect would have
quickly removed the smaller collisional remnants. In simulations by
\cite{Vokrouhlickyetal00} all Vulcanoids 1~km in diameter and smaller were
removed from the Vulcanoid zone by Yarkovsky drift, with typical lifetimes of
100~Myr for 0.1~km objects and 500~Myr for 1~km objects. Even objects up to
10~km in size showed strong depletion (80\%--100\%) over the age of the solar
system. By removing the small body population, the Yarkovsky effect might
greatly reduce the efficiency of collisional erosion in the Vulcanoid zone. To
date, there have been no studies combining the effects of collisional
evolution and Yarkovsky drift in the Vulcanoid zone.

If we assume that the present-day Vulcanoid population both exists and obeys a
steady-state collisional fragmentation size distribution, $dn=C*r^{-3.5}dr$
\citep{Dohnanyi1969, Williams:wetherill1994}, then our $3\sigma$ upper limit
implies that no more than 76 primordial Vulcanoids remain in the Vulcanoid
zone--the last remnants of a much larger initial population. Such Vulcanoids
would all be larger than 1~km in diameter, as primordial objects smaller than
this will have been removed from the Vulcanoid zone by Yarkovsky
drift. However, since the combination of collisional evolution and Yarkovsky
drift could result in either the complete depletion of a primordial Vulcanoid
population or something less than that, our upper limits cannot be
extrapolated to place any meaningful constraints on a hypothetical primordial
Vulcanoid population.

Finally, we note that if a Vulcanoid is subsequently discovered (hopefully
below our detection limits) it is possible that it may be a scattered
Near-Earth Object (NEO) rather than a primordial Vulcanoid. Recent modeling of
the orbital distribution of NEOs by \cite{Greenstreetetal12} suggests that up
to 0.006\% of the NEO population might end up in orbits with an aphelion
interior to Mercury's aphelion. Given that limit, an NEO interloper might have
an absolute magnitude as low as $H \simeq 21$, corresponding to a diameter
between 170-370~m, more than an order of magnitude smaller than the detection
limits of our search.


\acknowledgments We kindly thank D. Bewsher for providing the instrument
response curves for STEREO HI-1, G.~M. Holsclaw for providing the absolute
reflectance of Mercury as measured by MASCS/MDIS, T.~V. Spahr for initial
orbital solutions for the object that turned out to be C/2008 Y12 (SOHO), and
K. Battams for discussions about C/2008 Y12 (SOHO) and its orbit. Support for
this work was provided by NASA Planetary Geology and Geophysics program
through grant NNX09AD65G.

\appendix

\section{Statistical Tests}

To determine if there were any statistically significant deviations in the
detected and/or non-detected populations of synthetic Vulcanoids, we conducted
a series of statistical tests. The first test was to divide the range of
possible values for each orbital parameter into four, equally-spaced bins. If
the parameter values are randomly distributed, then the expectation is that
1/4 of the total population should have parameter values in the range covered
by each bin. The cumulative binomial distribution can be used to determine the
significance of any deviation from this expected value. The largest deviation
from a random uniform distribution occurred in the non-detected population:
the fourth semi-major axis bin had 17 objects, compared to an expectation
value of 10.75--an event with only a 2.5\% probability. While this may seem
marginally significant, given that we conducted 24 separate tests, we would
expect one of them to have as large a deviation nearly 46\% of the time. Thus,
we conclude there were no statistically significant deviations from a random
normal distribution.

We also performed one- and two-sample Kolmogorov-Smirnoff (K-S) tests on the
distributions of orbital parameters for both the detected and undetected
populations. A one-sample K-S test compares the empirical distribution
function of a sample population with the cumulative distribution function of a
reference distribution (in this case, a random uniform distribution) and
evaluates the significance of any deviations. None of the empirical
distribution functions of the six orbital parameters for either of the two
populations was significantly different from a random uniform
distribution. Similarly, a two-sample K-S test compares the empirical
distribution functions of two sample populations to determine if they are
consistent with being derived from the same parent distribution. We found no
statistically significant difference between the parent distributions of the
detected and non-detected populations for any of the six orbital parameters.

Finally, we examined each of the 21 combination of the six orbital parameters
and Vulcanoid diameter for both possible correlations. Both the Pearson linear
correlation coefficient (which measures how well two variables are linearly
related) and the Spearman rank correlation coefficient (which measures how
well the relationship between the two variables can be described using a
monotonic function) exhibited near-zero values for all parameter combinations.





\begin{deluxetable}{rrrrrrrrr}
  \tabletypesize{\small}
  \tablecolumns{8}
  \tablewidth{0pc}
  \tablecaption{Synthetic Objects Detected by Observer\label{deteff_table}}
  \tablehead{
    \colhead{Diameter (km)\tablenotemark{a}} & \colhead{X$_{obs 1}$} & 
    \colhead{X$_{obs 2}$} & \colhead{X$_{obs 3}$} & \colhead{X$_{obs 4}$} &
    \colhead{N} & \colhead{$\epsilon_{avg}$\tablenotemark{b}} &
    \colhead{P$_{det}$} & \colhead{P$^\prime_{det}$\tablenotemark{c}}}
  \startdata
   0.7--1.0    &   0 & 0 & 0 & 0 &   4 &  0.000$^{+0.065}$ &  0.000 & 0.517\tablenotemark{d}\\
   1.0--1.4    &   0 & 0 & 0 & 0 &   8 &  0.000$^{+0.034}$ &  0.000 & 0.344\tablenotemark{d}\\
   1.4--2.0    &   0 & 0 & 0 & 0 & 10 &  0.000$^{+0.028}$ &  0.000 & 0.294\tablenotemark{d} \\
   2.0--2.8    &   0 & 0 & 0 & 0 &   8 &  0.000$^{+0.034}$ &  0.000 & 0.344\tablenotemark{d} \\
   2.8--4.0    &   0 & 0 & 0 & 0 &   8 &  0.000$^{+0.034}$ &  0.000 & 0.344\tablenotemark{d} \\
   4.0--5.7    &   3 & 1 & 0 & 0 &   8 &  0.125$^{+0.065}_{-0.051}$ & 0.453 & 0.611\tablenotemark{d} \\
   5.7--8.0    & 10 & 9 & 8 & 5 & 10 &  0.800$^{+0.057}_{-0.067}$ &  1.000 & 0.998 \\
   8.0--11.3  &   8 & 8 & 8 & 8 &   8 &  1.000$_{-0.034}$ &  1.000 &  1.000\\
  11.3--16.0 &   8 & 8 & 8 & 8 &   8 &  1.000$_{-0.034}$ &  1.000 &  1.000\\
  16.0--22.6 &   8 & 8 & 8 & 8 &   8 &  1.000$_{-0.034}$ &  1.000 &  1.000\\
  22.6--32.0 &   9 & 9 & 9 & 9 &   9 &  1.000$_{-0.031}$ &  1.000 &  1.000\\
  32.0--45.3 &   8 & 8 & 8 & 8 &   8 &  1.000$_{-0.034}$ &  1.000 &  1.000\\
  45.3--64.0 &   4 & 4 & 4 & 4 &   4 &  1.000$_{-0.065}$ &  1.000 &  0.999 \\
\enddata
\tablenotetext{a}{Diameter assumes a Vulcanoid R-band albedo of 0.05. For other assumed
  values of the albedo, d$\propto$ p$_R^{-1/2}$.}
\tablenotetext{b}{Detection efficiency, averaged over the four observers, with
  1$\sigma$ (68.3\%) confidence intervals.}  
\tablenotetext{c}{Probability of at least one of the four observers detecting
  a single object, given statistical uncertainties associated with X$_{obs}$
  detections of N objects for each observer.}
\tablenotetext{d}{When X$_{obs}$=0 for at least one observer, P$^\prime_{det}$
  is likely to significantly overstate the detection probability. This results
  from the fact that given the small value of N, a non-zero value for
  $\epsilon_{obs}$ cannot be excluded with a high degree of confidence, even
  though in all likelihood, $\epsilon\approx0$}
\end{deluxetable}

\begin{deluxetable}{lllllll}
  \tabletypesize{\small}
  \tablecolumns{7}
  \tablewidth{0pc}
  \tablecaption{Orbital Elements for C/2008 Y12 (SOHO)\label{elements_table}}
  \tablehead{
    \colhead{Orbit Solution} & \colhead{q (AU)} & \colhead{e} & 
    \colhead{i ($^\circ$)} & \colhead{$\Omega$ ($^\circ$)} & 
    \colhead{$\omega$ ($^\circ$)} & \colhead{T$_0$ (UTC)}}
  \startdata
  \cite{MPEC2009F17} & 0.0533       &  1.0           & 154.34      &   35.03 & 140.06 & 2008 Dec.  22.65   \\
  \cite{MPEC2012B23} & 0.0658983 &  1.0           &  23.39475 & 312.10406 & 145.62274 & 2008 Dec. 22.60543 \\
  This work                 & 0.0659134 &  0.970428 &  23.38917 & 312.10392 & 146.56106 & 2008 Dec. 22.62134 \\
\enddata
\end{deluxetable}

\end{document}